\documentclass[12pt]{article}
\usepackage[a4paper, total={7in, 10in}]{geometry}
\usepackage[parfill]{parskip}
\usepackage{physics, tensor, float, subcaption}
\usepackage{graphicx}
\graphicspath{ {Plots/} }
\usepackage{jhep-mod}
\usepackage{bm}
\usepackage{soul}
\usepackage{amssymb,amsmath,amsthm}
\usepackage{mathrsfs}
\usepackage[utf8]{inputenc}
\usepackage{enumerate}
\usepackage{bigints}
\usepackage{xcolor}
\usepackage{appendix}
\usepackage{graphicx}
\usepackage{float}
\usepackage{tikz}
\usepackage{setspace}
\usepackage{cancel}
\usepackage{array}
\usepackage{tabulary}
\usepackage{doi}
\definecolor{purple}{rgb}{1,0,1}
\definecolor{lime}{HTML}{A6CE39} 

\definecolor{lime}{HTML}{A6CE39}
\newcommand{\orcidicon}{%
	\begin{tikzpicture}
	\draw[lime, fill=lime] (0,0) 
		circle [radius=0.16] 
		node[white] {{\fontfamily{qag}\selectfont \tiny ID}};
	\draw[white, fill=white] (-0.0625,0.095) 
		circle [radius=0.007];
	\end{tikzpicture}
	\hspace{-5mm}
}
\newcommand\orcidJosh{{\href{https://orcid.org/0000-0003-1200-7261}{\orcidicon}}}
\newcommand\orcidRudeep{{\href{https://orcid.org/0009-0002-0162-562X}{\orcidicon}}}
\newcommand\orcidMatt{{\href{https://orcid.org/0000-0003-1088-6485}{\orcidicon}}}

\newcommand{\be}{\begin{equation}gin{equation}}
\newcommand{\ee}{\end{equation}}

\begin{document}
\newcommand{\arXiv}[1]{arXiv:\href{https://arxiv.org/abs/#1}{\color{blue}#1}}


\title{\vspace{-25pt}\huge{
Defect wormholes are defective 
}}


\author{
\Large
Joshua Baines\!\orcidJosh, Rudeep Gaur\!\orcidRudeep,  {\sf  and} Matt Visser\!\orcidMatt}
\affiliation{School of Mathematics and Statistics, Victoria University of Wellington, 
\\
\null\qquad PO Box 600, Wellington 6140, New Zealand.}
\emailAdd{joshua.baines@sms.vuw.ac.nz}
\emailAdd{rudeep.gaur@sms.vuw.ac.nz}
\emailAdd{matt.visser@sms.vuw.ac.nz}
\renewcommand{\arXiv}[1]{arXiv:\href{https://arxiv.org/abs/#1}{\color{blue}#1}}
\def\L{{\mathcal{L}}}

\abstract{
\vspace{1em}

The various ``defect wormholes'' developed by Klinkhamer have recently attracted considerable attention --- especially in view of the fact that the simplest example, the so-called ``vacuum defect wormhole'', was claimed to be an everywhere-vacuum everywhere-Ricci-flat exact solution to the Einstein equations. 
This claim has been conclusively refuted by Feng, and in the current article we take a deeper look at exactly what goes wrong.
The central issue is this: Although Klinkhamer's specific representation of the metric $g_{ab}$ is smooth ($C^\infty$) his inverse metric $g^{ab}$  is not even everywhere continuous ($C^0$), being undefined at the wormhole throat. 
This situation implies that one should very carefully investigate curvature tensors at the throat using the Israel--Lanczos--Sen thin-shell formalism. 
Doing so reveals the presence of a delta-function energy-condition-violating thin shell of matter at the wormhole throat. 
The ``defect wormholes'' are thus revealed to be quite ordinary ``cut-and-paste'' thin-shell wormholes, but represented in a coordinate system which is unfortunately pathological at exactly the same place that all the interesting physics occurs. 
{To help clarify the situation, we shall focus on the behaviour of suitable coordinate invariants---the Ricci scalar, the eigenvalues of the mixed $R^a{}_b$ Ricci tensor, and the eigenvalues of the mixed $R^{ab}{}_{cd}$ Riemann tensor. }

\bigskip
\noindent
{\sc Date:} 31 August 2023; 17 October 2023; \LaTeX-ed \today

\bigskip
\noindent{\sc Keywords}: Wormholes,  thin-shell construction, energy conditions.

\bigskip
\noindent{\sc Published as}: Universe \textbf{9} (2023) 452.

\bigskip
\noindent{\sc PhySH:} 
Gravitation
}

\maketitle
\def\tr{{\mathrm{tr}}}
\def\diag{{\mathrm{diag}}}
\def\cof{{\mathrm{cof}}}
\def\pdet{{\mathrm{pdet}}}
\def\QED{ {\hfill$\Box$\hspace{-25pt}  }}
\def\d{{\mathrm{d}}}
\def\sign{\hbox{sign}}

\parindent0pt
\parskip7pt

\clearpage
\null
\vspace{-75pt}
\section{Introduction}

The so-called ``defect wormholes'' developed by Klinkhamer in references~\cite{Klinkhamer:2022,Klinkhamer:2023a,Klinkhamer:2023b,Klinkhamer:2023c} have recently attracted considerable attention~\cite{Wang:2023,Ahmed:2023a,Ahmed:2023b,Ahmed:2023c,Ahmed:2023d}, despite being in serious conflict with over 35 years of Lorentzian wormhole physics~\cite{Morris:1988a,Morris:1988b}. 
Unfortunately, it is our melancholy duty to confirm and extend the severe technical criticisms presented by Feng in reference~\cite{Feng:2023}. 

The basic issue is that Klinkhamer's coordinate choice is pathological exactly on the wormhole throat:
The metric $g_{ab}$ is smooth ($C^\infty$) but the inverse metric $g^{ab}$  is not even everywhere continuous ($C^0$), being undefined at the wormhole throat itself. 
This pathology has the effect of hiding the existence of a thin-shell delta-function layer of energy-condition-violating stress-energy at the wormhole throat.
\begin{itemize}

\item 
We shall first, using Klinkhamer's pathological coordinates, verify the existence of a discontinuity in the extrinsic curvature of the spherically symmetric (2+1) ``constant radius'' hyper-surfaces 
at the wormhole throat. 
We note that this discontinuity in the extrinsic curvature occurs at exactly the same place that the underlying coordinate system becomes pathological, which is why symbolic manipulation software, or indeed na\"ive computations ``by hand'',  often lead to misleading results --- this is a situation where careful analytic insight is called for.
We shall furthermore relate this discontinuity in the extrinsic curvature to the de-focussing properties of the wormhole throat and thence to violations of the null curvature condition.

\item
We shall then set up a more reasonable ``proper distance'' coordinate chart, effectively amounting to the use of Gaussian normal coordinates, and
invoke the Israel--Lanczos--Sen thin-shell formalism~\cite{Israel, Lanczos:1922,Lanczos:1924, Sen}, 
early versions of which are now 99 years old,    to get a better grasp on the physics at the wormhole throat.
In this improved coordinate chart  the metric $g_{ab}$  and inverse metric $g^{ab}$ are both at least $C^0$, and are almost everywhere $C^1$. (That is, $C^{1-}\,$, meaning differentiable but not continuously so, with a discontinuity in the Christoffel symbols  at the wormhole throat.)

\item
We then explicitly calculate the thin-shell delta-function contributions to the curvature tensors, (and thence, invoking the Einstein equations, to the stress-energy). 
 
  \item
  {To help make the analysis robust we shall focus on the behaviour of suitable coordinate invariants --- the Ricci scalar, the eigenvalues of the mixed $R^a{}_b$ Ricci tensor, and the eigenvalues of the mixed $R^{ab}{}_{cd}$ Riemann tensor.  (The more usual polynomial curvature invariants are not useful in that they correspond to ill-defined squares and higher powers  of delta-functions.)}

\end{itemize}\enlargethispage{50pt}
Overall this implies that that the “vacuum defect” wormholes are really just quite standard 
examples of thin-shell cut-and-paste wormholes in disguise~\cite{examples,surgery}, these thin-shell cut-and-paste wormholes dating back at least to 1989.
(For more related discussion see references~\cite{book, Poisson:1995,Musgrave:1995} and more recently~\cite{Eiroa:2003, Lobo:2003, Lobo:2004, Lobo:2004b, Lobo:2005b,toolkit, small, Kar:2004, Sharif:2016}.)
We find, as expected,  the usual delta-function   
layer of energy-condition-violating stress-energy at the wormhole throat~\cite{Morris:1988a,Morris:1988b,examples,surgery,book}. 

\section{Vacuum defect wormhole}

This simplest ``vacuum defect wormhole'' is an example which already contains all of the important physics issues, 
and will serve as a template for the various variant defect wormholes subsequently considered. 

\subsection{Pathological coordinates}

Klinkhamer's original ``vacuum defect wormhole'' corresponds to the line-element~\cite{Klinkhamer:2022}
\begin{equation}
\label{E:line}
 \d s^2 = -\d t^2 + {\xi^2\,  \d\xi^2\over \lambda^2+\xi^2} + (\lambda^2+\xi^2) \d \Omega^2; \qquad \xi \in (-\infty,+\infty).
\end{equation}
Yes, this line element can be given a wormhole interpretation:  There is a wormhole throat, of size $\lambda$ and area $4\pi\lambda^2$, located at $\xi=0$, 
and \emph{na\"ively} the Riemann tensor is zero everywhere. (The implied existence of a  \emph{na\"ively} everywhere-Riemann-flat wormhole should really be cause to make one stop, re-assess, and carefully reflect on the underlying physics.) 
A more correct statement is that the Riemann tensor is zero almost everywhere, \emph{except} at the throat $\xi=0$, where {the \emph{na\"ive} coordinate system breaks down and a}
more careful analysis is required. 
A quick way to see that some care is needed at the wormhole throat is to notice that in Klinkhamer's coordinates
\begin{equation}
g^{\xi\xi} =  1 + {\lambda^2\over\xi^2};
\end{equation}
so the inverse metric is ill-defined at the wormhole throat. (The inverse metric components are not even $C^0$.)
This is considerably worse than what we know happens at the Schwarzschild horizon in standard coordinates --- since in the current setup the metric determinant itself is actually zero --- $\det(g_{ab})\propto \xi^2 (\lambda^2+\xi^2) \to 0$. For the inverse metric at the throat we have $\det(g^{ab})\to\infty$.
Furthermore we note
\begin{equation}
\Gamma^\xi{}_{\xi\xi} = {\lambda^2\over\xi(\lambda^2+\xi^2)}; \qquad 
\Gamma^\xi{}_{\theta\theta} = -{\lambda^2+\xi^2\over\xi}; \qquad 
\Gamma^\xi{}_{\phi\phi} = -{\lambda^2+\xi^2\over\xi}.
\end{equation}
That is, in these particular coordinates,  three of the Christoffel symbols of the 2$^{nd}$  kind exhibit infinite discontinuities at the wormhole throat.
(The other Christoffel symbols are smooth.)
These observations all point to the fact that one needs to be extremely careful when analyzing what is happening at the wormhole throat $\xi=0$,  which is exactly where the present coordinate system is problematic.

\enlargethispage{40pt}
Let us now  calculate the extrinsic curvature of the constant-$\xi$ hyper-surfaces.  

In the current $\xi$ coordinate system $(t,\xi,\theta,\phi)$ the unit normal $n^a$ to the constant-$\xi$ hyper-surfaces is
\begin{equation}
n_a = {|\xi|\over \sqrt{\lambda^2+\xi^2}} \;(0,1,0,0)_a ;  \qquad 
n^a = {\sqrt{\lambda^2+\xi^2}\over |\xi|}\;(0,1,0,0)^a.
\end{equation}
The normal has been carefully constructed to always point from one asymptotic region into the other, specifically, in the direction of increasing $\xi$. (This step is essential to giving the line element (\ref{E:line}) a wormhole interpretation.)
Then in this coordinate system the extrinsic curvature of the constant-$\xi$ hyper-surfaces is

\begin{equation}
K_{ab}= -n_{(a;b)} = {1\over2} \L_n g_{ab} 
= {1\over2}\left[ n^c\partial_c g_{ab} -  (\partial_a n^c) g_{cb} - (\partial_b n^c) g_{ac}\right].
\end{equation}
Thence in this particular situation
\begin{equation}
K_{ab}  = {1\over2} n^c\partial_c [g_{ab} - n_a n_b],
\end{equation}
leading to
\begin{equation}
K_{ab} =  \sign(\xi)\; \sqrt{\lambda^2 +\xi^2} 
 \left[\begin{array}{cccc}
0&0&0&0\\{}0&0&0&0\\{}0&0&1&0\\{}0&0&0&\sin^2\theta 
\end{array}\right].
\end{equation}
We can also write this as 
\begin{equation}
K^a{}_{b} =  {\sign(\xi)\over \sqrt{\lambda^2 +\xi^2} } 
\left[\begin{array}{cccc}
0&0&0&0\\{}0&0&0&0\\{}0&0&1&0\\{}0&0&0&1
\end{array}\right],
\end{equation}
implying a step-function discontinuity in the extrinsic curvature at the wormhole throat, 
\begin{equation}
\Delta K^a{}_b \;=\; \left.K^a{}_b\right|_{\xi=0^+}- \left.K^a{}_b\right|_{\xi=0^-} \;= \;{2\over  \lambda} \left[\begin{array}{cccc}
0&0&0&0\\{}0&0&0&0\\{}0&0&1&0\\{}0&0&0&1 
\end{array}\right],
\end{equation}
and so implying, (via the Israel--Lanczos--Sen thin-shell formalism~\cite{Israel, Lanczos:1924, Lanczos:1922, Sen}), a delta-function contribution to the Riemann, Weyl, Ricci, and Einstein tensors at the wormhole throat. 
\enlargethispage{30pt}

The necessity for such thin-shell delta-function contributions to the curvature can be deduced on quite general grounds: As already noted in~\cite{Morris:1988a,Morris:1988b}, incoming radial null curves in one universe become outgoing radial null curves in the other universe, so the wormhole throat acts as a diverging lens. But in this ``vacuum defect'' model there is certainly  no spacetime curvature away from the throat, so this defocussing must be accomplished by curvature concentrated \emph{on the throat itself}. Indeed, appeal to the Raychaudhuri equation~\cite{Raychaudhuri} or its variants~\cite{Dadhich:2005,Kar:2006,Abreu:2010}  implies violation of the Ricci null curvature condition on the throat: $R_{ab} \,k^a \,k^b < 0$ for radial null vectors. Invoking the standard Einstein equations converts this to a violation of the null energy condition: $T_{ab} \,k^a\,k^b <0$ on the throat.

But to fully complete the job, and quantitatively fix the overall normalization of these thin-shell contributions, one has to introduce a Gaussian normal coordinate patch straddling the wormhole throat, a task to which we now turn.

\subsection{Non-pathological coordinates: Gaussian normal coordinates}

To get a better grasp as to what is going on, define the (signed) proper distance coordinate measuring physical distance from the wormhole throat:
\begin{equation}
\ell(\xi) = \sign(\xi) \int_0^\xi {\xi \d\xi\over \sqrt{\lambda^2+\xi^2}} = \sign(\xi) \left[ \sqrt{\lambda^2+\xi^2} - \lambda\right]; 
\qquad \ell \in (-\infty,+\infty).
\end{equation}
Then we see
\begin{equation}
\label{E:proper}
\lambda+|\ell|= \sqrt{\lambda^2+\xi^2}; \qquad\hbox{so}\qquad
(\lambda+|\ell|)^2 = {\lambda^2+\xi^2}.
\end{equation}
In these proper distance coordinates the line element becomes 
\begin{equation}
\label{E:proper-line-element}
 \d s^2 = -\d t^2 + \d\ell^2 + \left(\lambda+ |\ell|\right)^2 \d \Omega^2; \qquad \ell \in (-\infty,+\infty).
\end{equation}
This line element is $C^0$ everywhere and $C^1$ almost everywhere, but not $C^1$ at the wormhole throat.
Indeed both the metric $g_{ab}$ and its inverse $g^{ab}$ are now $C^0$ (but not quite $C^1$ at the wormhole throat).
The notation $C^{1-}$ denotes this situation, piecewise differentiable with at most isolated step-function discontinuities. 
Thence in particular the Christoffel symbols will at worst contain step-function discontinuities.
So the curvature tensors will at worst contain delta-function contributions.
In the $\ell$ coordinate system $(t,\ell,\theta,\phi)$ the unit normal to the constant-$\ell$ hyper-surfaces is trivially and quite simply
\begin{equation}
n_a = (0,1,0,0)_a ;  \qquad n^a = (0,1,0,0)^a
\end{equation}
In this $\ell$ coordinate system the computation of the extrinsic curvature of the constant-$\ell$ hyper-surfaces is slightly simpler than in the $\xi$ coordinate system. Indeed
\begin{equation}
\label{E:extrinsic-1}
K^a{}_b =  {1\over2} g^{ac} \;(n^d \partial_d) \, g_{cb} =  {1\over2} g^{ac} \;\partial_\ell \, g_{cb} 
=
{\sign(\ell)\over |\ell|+ \lambda} \left[\begin{array}{cccc}
0&0&0&0\\{}0&0&0&0\\{}0&0&1&0\\{}0&0&0&1 
\end{array}\right].
\end{equation}
This again implies the same step-function discontinuity in the extrinsic curvature at the wormhole throat, 
\begin{equation}
\Delta K^a{}_b \;=\; \left.K^a{}_b\right|_{\ell=0^+}- \left.K^a{}_b\right|_{\ell=0^-} \;= {2\over  \lambda} \left[\begin{array}{cccc}
0&0&0&0\\{}0&0&0&0\\{}0&0&1&0\\{}0&0&0&1 
\end{array}\right].
\end{equation}

This line element is now easily recognized as a ``cut and paste'' traversable wormhole, as discussed in references~\cite{examples,surgery} and ~\cite{book}.
One simply takes two copies of Minkowski space, excises a timelike tube of radius $\lambda$ from each copy, and identifies the (2+1) dimensional surfaces of the two timelike tubes.  The two open regions $\ell\in(0,+\infty)$ and $\ell\in(-\infty,0)$ each represent open subsets of flat Minkowski space, with all of the interesting wormhole physics confined to the throat at $\ell=0$. 

\subsection{Christoffel symbols}

In view of the above, in this proper distance Gaussian normal coordinate system we can write
\begin{equation}
g_{ab,c} = {\sign(\ell)\over2} \; \Delta K_{ab} \;n_c  + \hbox{(something smooth)}.
\end{equation}
This manifestly has the right symmetries and the right discontinuity.
Thence for the Christoffel symbols
\begin{equation}
\Gamma^a{}_{bc} = {1\over2} {\sign(\ell)} \; \left[\Delta K^a{}_b\; n_c + \Delta K^a{}_c \; n_b   - n^a \Delta K_{bc}  \right]  + \hbox{(something smooth)}.
\end{equation}
Thence for the derivatives of the Christoffel symbols
\begin{eqnarray}
\Gamma^a{}_{bc,d} &=& {\delta(\ell)} \; \left[\Delta K^a{}_b\; n_c\; n_d + \Delta K^a{}_c \; n_b \; n_d  - n^a \Delta K_{bc} \; n_d  \right] 
\nonumber\\
&&\qquad  + \hbox{(something piecewise smooth)}.
\end{eqnarray}

\subsection{Curvature tensors}
\enlargethispage{40pt}

From the discussion above, we already see that for the Riemann tensor we have
\begin{eqnarray}
R^a{}_{bcd} &=& - {\delta(\ell)} \; \left[ \Delta K^a{}_c \; n_b \; n_d -  \Delta K^a{}_d \; n_b \; n_c
+ n^a \Delta K_{bd} \; n_c- n^a \Delta K_{bc} \; n_d  \right] 
\nonumber\\
&&\qquad  + \hbox{(something piecewise smooth)}.
\end{eqnarray}
See also the related discussion in reference~\cite{book}. 
But, since we have already seen that for the ``vacuum defect'' wormhole (\ref{E:proper-line-element}) the Riemann tensor is zero everywhere away from the wormhole throat, this actually implies the considerably stronger statement
\begin{eqnarray}
R^a{}_{bcd} &=& - {\delta(\ell)} \; \left[ \Delta K^a{}_c \; n_b \; n_d -  \Delta K^a{}_d \; n_b \; n_c
+ n^a \Delta K_{bd} \; n_c- n^a \Delta K_{bc} \; n_d  \right].
\end{eqnarray}
Equivalently
\begin{eqnarray}
R_{abcd} &=& - {\delta(\ell)} \; \left[ \Delta K_{ac} \; n_b \; n_d -  \Delta K_{ad} \; n_b \; n_c
+ \Delta K_{bd} \; n_a\; n_c-  \Delta K_{bc} \; n_a \; n_d  \right].
\end{eqnarray}

Thence for the Ricci tensor 
\begin{eqnarray}
R_{ab} &=& - {\delta(\ell)} \; \left[  \Delta K_{ab} + \Delta K \; n_a \; n_b   \right],\qquad
\end{eqnarray}
{and for the Ricci scalar
\begin{eqnarray}
R &=& -2 {\delta(\ell)} \; \left[ \Delta K  \right].
\end{eqnarray}
(In particular, the Ricci scalar, a coordinate invariant, explicitly exhibits a delta-function contribution at the wormhole throat.)
}

Thence for the Einstein tensor
\begin{eqnarray}
G_{ab} &=& - {\delta(\ell)} \; \left[ 
 \Delta K_{ab} - \Delta K \; (g_{ab} -n_a \; n_b )  \right].
\end{eqnarray}

Finally for the Weyl tensor it is useful to introduce the quantities
\begin{equation}
\widetilde{\Delta K}_{ab} = \Delta K_{ab} - {1\over3} \Delta K \;(g_{ab}- n_a n_b),
\end{equation}
which represents the traceless part of the jump in extrinsic curvature,
and
\begin{equation}
\tilde g_{ab} =  g_{ab}-2 n_a n_b,
\end{equation}
which reverses the radial part of the metric.
A quick calculation then yields (see for instance~\cite{book}):
\begin{equation}
C_{abcd} = 
 {1\over2} {\delta(\ell)} \; \left[ \widetilde{\Delta K}_{ac} \; \tilde g_{bd} - \widetilde{\Delta K}_{ad} \; \tilde g_{bc} 
+ \widetilde{\Delta K}_{bd} \; \tilde g_{ac} -  \widetilde{\Delta K_{bc}} \; \tilde g_{ad} \right].
\end{equation}
It is easy to check that the Weyl tensor is traceless on all pairs of indices.

{
\subsection{Other coordinate invariants}
In the absence of delta-function contributions to the curvature, one typically  works with polynomial curvature invariants 
such as $R_a{}^bR_b{}^a$,\; $R_a{}^bR_b{}^c R_c{}^a$,\; \emph{etc.}, or the Kretschman scalar $R_{abcd} R^{abcd}$, the Weyl scalar 
$C_{abcd} C^{abcd}$, or their generalizations.
In the presence of delta-function contributions to the curvature, since $\delta(\eta)^n$ is ill-defined for $n\geq 2$, the usual polynomial curvature invariants are ill-defined on the support of the delta-function, and a more subtle approach is called for.

\clearpage
A suitable class of coordinate invariants which one might consider are the eigenvalues of the mixed Ricci tensor $R^a{}_b$. One defines
\begin{equation}
R^a{}_b \; V^b = \lambda_{Ricci} \;V^a.
\end{equation}
The eigenvalues $\lambda_{Ricci}$ are then coordinate invariants, whereas the eigenvectors $V^a$ are coordinate \emph{covariant}. 
These eigenvalues are closely related to (and a simplification of) the Segre classification of the Ricci tensor~\cite{Segre, Plebanski, Santos:2005}.
A formally identical construction applies to the Einstein tensor, or indeed any symmetric rank 2 tensor.
In the current situation these eigenvalues will at worst contain one delta-function, and so will be well defined. 

Similarly we can consider the eigenvalues of the mixed Riemann tensor $R^{ab}{}_{cd}$ which can be viewed as a mapping of (contravariant) 2-forms to (contravariant) 2-forms:
\begin{equation}
R^{ab}{}_{cd} \; \omega^{cd} = \lambda_{Riemann} \;\omega^{ab}.
\end{equation}
The eigenvalues $\lambda_{Riemann}$ are then coordinate invariants, while the eigen-tensors $\omega^{ab}$ are coordinate \emph{covariant}. 
These eigenvalues  are closely related to (and a simplification of) the Petrov classification~\cite{Petrov,Stephani:2003}.
A formally identical construction applies to the Weyl tensor, or indeed to any rank 4 tensor that has the same algebraic symmetries as the Riemann/Weyl tensors.
In the current situation these eigenvalues will at worst contain one delta-function, and so will be well defined. 

We shall explicitly exhibit some of these coordinate invariant eigenvalues in the discussion below.
}

\subsection{Orthonormal tetrad basis}

If we work in an orthonormal tetrad basis then explicitly
\begin{equation}
\Delta K_{\hat a\hat b} =
{2\over  \lambda} \left[\begin{array}{cccc}
0&0&0&0\\{}0&0&0&0\\{}0&0&1&0\\{}0&0&0&1 
\end{array}\right]; 
\qquad \hbox{and} \qquad
\Delta K = {4\over\lambda}.
\end{equation}
Thence
\begin{equation}
R_{\hat a\hat b} =
-{2\over  \lambda} \left[\begin{array}{cccc}
0&0&0&0\\{}0&2&0&0\\{}0&0&1&0\\{}0&0&0&1 
\end{array}\right]  \delta(\ell);
\qquad
R = -{8\over\lambda}\delta(\ell);
\qquad
G_{\hat a\hat b} =
{2\over  \lambda} \left[\begin{array}{cccc}
-2&0&0&0\\{}0&0&0&0\\{}0&0&1&0\\{}0&0&0&1 
\end{array}\right]  \delta(\ell).
\end{equation}
Notice the explicit on-throat violation of the null curvature condition.

Furthermore, since
\begin{equation}
R^{\hat a}{}_{\hat b} =
-{2\over  \lambda} \left[\begin{array}{cccc}
0&0&0&0\\{}0&2&0&0\\{}0&0&1&0\\{}0&0&0&1 
\end{array}\right]  \delta(\ell),
\end{equation}
is diagonal we can immediately read off the four coordinate invariant eigenvalues of the mixed Ricci tensor as
\begin{equation}
\lambda\left(R^a{}_b\right) = \lambda\left(R^{\hat a}{}_{\hat b}\right) =
 \left\{ 0, -{4\delta(\ell)\over\lambda},-{2\delta(\ell)\over\lambda},-{2\delta(\ell)\over\lambda} \right\}.
\end{equation}

In the orthonormal basis (up to permutation symmetries)  the Riemann tensor has only two non-zero components:
\begin{equation}
R_{\hat\ell\hat\theta\hat\ell\hat\theta} = R_{\hat\ell\hat\phi\hat\ell\hat\phi} = -{2\over\lambda} \; \delta(\ell).
\end{equation}
Thence
\begin{equation}
R^{\hat\ell\hat\theta}{}_{\hat\ell\hat\theta} = R^{\hat\ell\hat\phi}{}_{\hat\ell\hat\phi} = -{2\over\lambda} \; \delta(\ell),
\end{equation}
and so we can read off the six coordinate invariant eigenvalues of the mixed Riemann tensor as
\begin{equation}
\lambda\left(R^{ab}{}_{cd}\right) = \lambda\left( R^{\hat\ell\hat\phi}{}_{\hat\ell\hat\phi} \right)=
\left\{  -{2\over\lambda} \; \delta(\ell) , \; -{2\over\lambda} \; \delta(\ell),\;  0,0,0,0\right\}.
\end{equation}

Finally for the Weyl tensor one has
\begin{equation}
C_{\hat t\hat \ell \hat t \hat \ell} = C_{\hat\theta\hat\phi\hat\theta\hat\phi} = {2\over3\lambda} \; \delta(\ell),
\end{equation}
and
\begin{equation}
C_{\hat t\hat \theta \hat t \hat \theta} = C_{\hat t\hat \phi \hat t \hat \phi} 
= C_{\hat\ell\hat\theta\hat\ell\hat\theta} = C_{\hat\ell\hat\phi\hat\ell\hat\phi} 
= -{1\over3\lambda} \; \delta(\ell).
\end{equation}
Thence
\begin{equation}
C^{\hat t\hat \ell}{}_{\hat t \hat \ell} = C^{\hat\theta\hat\phi}{}_{\hat\theta\hat\phi} = {2\over3\lambda} \; \delta(\ell),
\end{equation}
and
\begin{equation}
C^{\hat t\hat \theta}{}_{\hat t \hat \theta} = C^{\hat t\hat \phi}{}_ {\hat t \hat \phi} 
= C^{\hat\ell\hat\theta}{}_{\hat\ell\hat\theta} = C^{\hat\ell\hat\phi}{}_{\hat\ell\hat\phi} 
= -{1\over3\lambda} \; \delta(\ell),
\end{equation}
and so we can read off the six coordinate invariant eigenvalues of the mixed Weyl tensor as
\begin{equation}
\lambda\left(C^{ab}{}_{cd}\right) = \lambda\left( C^{\hat\ell\hat\phi}{}_{\hat\ell\hat\phi} \right)=
\left\{  {2\over3\lambda} \; \delta(\ell) , \; {2\over3\lambda} \; \delta(\ell),\; 
 - {1\over3\lambda} \delta(\ell), -{1\over3\lambda} \delta(\ell), -{1\over3\lambda} \delta(\ell), -{1\over3\lambda} \delta(\ell)\right\}.
\end{equation}\enlargethispage{30pt}
Note that all the curvature tensors, and their related coordinate invariant eigenvalues, are explicitly almost-everywhere zero --- except for a delta-function contribution located exactly on the wormhole throat.

\subsection{Stress-energy tensor}

Invoking the standard Einstein equations, we see that there is a delta-function contribution the stress-energy tensor, and the stress-energy tensor \emph{does} violate the energy conditions.
Specifically the surface energy density $\sigma$ and surface pressure $\wp$ on the wormhole throat are~\cite{examples,surgery,book}:
\begin{equation}
\sigma = -{1\over2\pi G_N \lambda}; \qquad\qquad \wp= +{1\over 4\pi G_N \lambda}.
\end{equation}
This manifestly violates the WEC, and indeed (since $\sigma+\wp < 0$) also violates the NEC, SEC, and DEC.
%
If one wishes, one can explicitly recast this in terms of a delta-function contribution to the stress-energy: 
\begin{equation}
T^{\hat a \hat b} = {1\over 4\pi G_N \lambda} \left[\begin{array}{cccc}
-2&0&0&0\\{}0&0&0&0\\{}0&0&1&0\\{}0&0&0&1 
\end{array}\right] \delta(\ell).
\end{equation}

 \section{More general defect wormhole}

Klinkhamer's (general)  ``defect wormhole'' corresponds to the somewhat more general line element
\begin{equation}
\label{E:line2}
 \d s^2 = -\d t^2 + {\xi^2 \d\xi^2\over \lambda^2+\xi^2} + (b_0^2+\xi^2) \d \Omega^2; \qquad \xi \in (-\infty,+\infty).
\end{equation}
Yes, this line element can again be given a wormhole interpretation: There is still a wormhole throat at $\xi=0$,
but now the quantities $b_0$ and $\lambda$ are distinct parameters.
This new spacetime is somewhat more general than the ``vacuum defect wormhole'' considered above; but no really new ideas are involved.
There is  still a delta-function contribution at the wormhole throat to both the curvature tensors and the stress-energy tensor, a contribution that is easy to miss if one is relying on symbolic algebra software, or  is na\"ive with computations carried out ``by hand''.

To see what is going on we again define
\begin{equation}
\ell(\xi) = \sign(\xi) \int_0^\xi {\xi \d\xi\over \sqrt{\lambda^2+\xi^2}} = \sign(\xi) \left[ \sqrt{\lambda^2+\xi^2} - \lambda\right]; 
\qquad \ell \in (-\infty,+\infty).
\end{equation}
Then $\ell(\xi)$ is the proper distance from the wormhole throat and we again see
\begin{equation}
\lambda+|\ell| = \sqrt{\lambda^2+\xi^2};
\qquad
(|\ell| + \lambda)^2 = {\lambda^2+\xi^2}.
\end{equation}

\clearpage
So in these proper distance coordinates the ``defect wormhole'' line-element (\ref{E:line2}) now becomes 
\begin{equation}
\label{E:proper2}
 \d s^2 = -\d t^2 + \d\ell^2 + \left\{ b_0^2-\lambda^2 + \left(|\ell|+\lambda\right)^2\right\} \d \Omega^2; \qquad \ell\in (-\infty,+\infty).
\end{equation}

This is again $C^0$ everywhere (but not $C^1$ at the wormhole throat).
There will again be step functions present in $\Gamma^a{}_{bc}$, delta-functions present in all of the curvature tensors,
and delta-functions present in the stress-energy.
There will now also be smooth non-zero contributions to these quantities away from the wormhole throat.

To proceed, one can first easily calculate the extrinsic curvatures of the  constant $\ell$ surfaces:
\begin{equation}
K^a{}_b = {\sign(\ell) \; (|\ell|+ \lambda) \over  b_0^2-\lambda^2 + \left(|\ell|+\lambda\right)^2} \left[\begin{array}{cccc}
0&0&0&0\\{}0&0&0&0\\{}0&0&1&0\\{}0&0&0&1 
\end{array}\right].
\end{equation}
Note the slightly different prefactor, but same basic structure, as in the ``vacuum defect'' wormhole of equations (\ref{E:proper-line-element}) and (\ref{E:extrinsic-1}).

This in turn implies a (slightly modified) step-function discontinuity in the extrinsic curvature at the wormhole throat, 
\begin{equation}
\Delta K^a{}_b \;=\; \left.K^a{}_b\right|_{\ell=0^+}- \left.K^a{}_b\right|_{\ell=0^-} \;= {2\lambda\over b_0^2} \left[\begin{array}{cccc}
0&0&0&0\\{}0&0&0&0\\{}0&0&1&0\\{}0&0&0&1 
\end{array}\right],
\end{equation}
and so (as per the previous discussion) a delta function contribution to the Riemann, Ricci, and Einstein tensors at the wormhole throat.

\subsection{Curvature tensors}

For the Riemann tensor we now have
\begin{eqnarray}
R^a{}_{bcd} &=& - {\delta(\ell)} \; \left[ \Delta K^a{}_c \; n_b \; n_d -  \Delta K^a{}_d \; n_b \; n_c
+ n^a \Delta K_{bd} \; n_c- n^a \Delta K_{bc} \; n_d  \right] 
\nonumber\\
&&\qquad\qquad  + \hbox{(something piecewise smooth)}.
\end{eqnarray}
But now we cannot simply discard the piecewise smooth contributions as they will generically be non-zero (if maybe relatively uninteresting).
Equivalently
\begin{eqnarray}
R_{abcd} &=& - {\delta(\ell)} \; \left[ \Delta K_{ac} \; n_b \; n_d -  \Delta K_{ad} \; n_b \; n_c
+ \Delta K_{bd} \; n_a\; n_c-  \Delta K_{bc} \; n_a \; n_d  \right] 
\nonumber\\
&&\qquad\qquad  + \hbox{(something piecewise smooth)},
\end{eqnarray}
{and
\begin{eqnarray}
R^{ab}{}_{cd} &=& - {\delta(\ell)} \; \left[ \Delta K^a{}_c \; n^b \; n_d -  \Delta K^a{}_d \; n^b \; n_c
+ n^a \Delta K^b{}_{d} \; n_c- n^a \Delta K^b{}_{c} \; n_d  \right] 
\nonumber\\
&&\qquad\qquad  + \hbox{(something piecewise smooth)}.
\end{eqnarray}
Thence the on-throat (coordinate invariant) eigenvalue structure will be (up to a constant multiplicative factor) identical to that for the ``vacuum defect'' wormhole. 
(There will now be additional off-throat piecewise smooth contributions to the eigenvalues, but they are not of central interest to the present discussion.) The key point is that there are still non-zero on-throat delta-function contributions to the coordinate invariant eigenvalues that closely mimic (up to a constant multiplicative factor) the behaviour we have already seen for the ``vacuum defect'' wormhole.
}

Furthermore for the Ricci tensor
\begin{eqnarray}
R_{ab} &=& - {\delta(\ell)} \; \left[  \Delta K_{ab} + \Delta K \; n_a \; n_b   \right]  
+ \hbox{(something piecewise smooth)},
\end{eqnarray}
and Ricci scalar
\begin{eqnarray}
R &=& -2  {\delta(\ell)} \; \left[ \Delta K  \right]  + \hbox{(something piecewise smooth)}.
\end{eqnarray}
Thence for the Einstein tensor
\begin{eqnarray}
G_{ab} &=& - {\delta(\ell)} \; \left[ 
 \Delta K_{ab} - \Delta K \; (g_{ab} -n_a \; n_b )  \right]  + \hbox{(something piecewise smooth)}.\qquad
\end{eqnarray}\enlargethispage{30pt}

Finally for the Weyl tensor
\begin{eqnarray}
C_{abcd} &=& 
 {1\over2} {\delta(\ell)} \; \left[ \widetilde{\Delta K}_{ac} \; \tilde g_{bd} - \widetilde{\Delta K}_{ad} \; \tilde g_{bc} 
+ \widetilde{\Delta K}_{bd} \; \tilde g_{ac} -  \widetilde{\Delta K_{bc}} \; \tilde g_{ad} \right]
\nonumber\\
&&\qquad\qquad  + \hbox{(something piecewise smooth)}.
\end{eqnarray}
As compared to the ``vacuum defect wormhole'' of (\ref{E:proper-line-element}) the only changes are the simple rescaling ${1\over\lambda}\to {\lambda\over b_0^2}$ and introduction of bulk piecewise smooth contributions away from the wormhole throat. 

It is easy to calculate the bulk piecewise smooth contributions, but they are relatively uninteresting, apart from checking that they vanish in the limit $b_0\to\lambda$.

\subsection{Orthonormal tetrad basis}

If we work in an orthonormal tetrad basis then 
\begin{equation}
\Delta K_{\hat a\hat b} =
{2\lambda\over b_0^2} \left[\begin{array}{cccc}
0&0&0&0\\{}0&0&0&0\\{}0&0&1&0\\{}0&0&0&1 
\end{array}\right]; 
\qquad\qquad
\Delta K = {4\lambda\over b_0^2}.
\end{equation}
Thence
\begin{equation}
R_{\hat a\hat b} =
-{2\lambda\over b_0^2}  \left[\begin{array}{cccc}
0&0&0&0\\{}0&2&0&0\\{}0&0&1&0\\{}0&0&0&1 
\end{array}\right] \delta(\ell) + \hbox{(something piecewise smooth)};
\end{equation}
\begin{equation}
R = -{8\lambda\over b_0^2}\delta(\ell) + \hbox{(something piecewise smooth)};
\end{equation}
\begin{equation}
G_{\hat a\hat b} =
{2\lambda\over b_0^2}  \left[\begin{array}{cccc}
-2&0&0&0\\{}0&0&0&0\\{}0&0&1&0\\{}0&0&0&1 
\end{array}\right] \delta(\ell) + \hbox{(something piecewise smooth)}.
\end{equation}

In the orthonormal basis (up to permutation symmetries)  the Riemann tensor has only two delta-function contributions:
\begin{equation}
R_{\hat\ell\hat\theta\hat\ell\hat\theta} = R_{\hat\ell\hat\phi\hat\ell\hat\phi} 
= -{2\lambda\over b_0^2} \; \delta(\ell) + \hbox{(something piecewise smooth)}.
\end{equation}
Finally for the only delta-function contributions to the Weyl tensor one has
\begin{equation}
C_{\hat t\hat \ell \hat t \hat \ell} = C_{\hat\theta\hat\phi\hat\theta\hat\phi} 
= {2\lambda\over3b_0^2} \; \delta(\ell) + \hbox{(something piecewise smooth)},
\end{equation}
and
\begin{equation}
C_{\hat t\hat \theta \hat t \hat \theta} = C_{\hat t \hat \phi \hat t \hat \phi} 
= C_{\hat\ell\hat\theta\hat\ell\hat\theta} = C_{\hat\ell\hat\phi\hat\ell\hat\phi} 
= -{\lambda\over3b_0^2} \; \delta(\ell) + \hbox{(something piecewise smooth)}.
\end{equation}

\subsection{Stress-energy tensor}

If one wishes, using the Einstein equations one can explicitly recast the stress-energy tensor in terms of a delta-function as 
\begin{equation}
T_{\hat a\hat b} =  {\lambda\over 4\pi G_N \; b_0^2} \left[\begin{array}{cccc}
-2&0&0&0\\{}0&0&0&0\\{}0&0&1&0\\{}0&0&0&1 
\end{array}\right] \delta(\ell)
+ \hbox{(something piecewise smooth)}.
\end{equation}
Regardless of  the behaviour of the smooth contribution to the stress-energy, the delta-function contribution manifestly violates the WEC, and indeed also violates the NEC, SEC, and DEC.

\clearpage
Specifically the surface energy density $\sigma$ and surface pressure $\wp$ on the wormhole throat are~\cite{examples,surgery,book}:
\begin{equation}
\sigma = -{\lambda\over2\pi G_N b_0^2}; \qquad\qquad \wp= +{\lambda\over 4\pi G_N b_0^2}.
\end{equation}
 The piecewise smooth contributions to the stress-energy are all proportional to $\lambda^2-b_0^2$ and so might or might not violate the energy conditions depending on the relative magnitudes of $\lambda$ and $b_0$. However in these ``defect wormhole'' models the on-throat violations of the energy conditions is completely generic and unavoidable.

\section{Discussion}

Overall, we see that the ``defect wormholes'' introduced by Klinkhamer do not actually represent new physics --- they are merely quite standard thin-shell ``cut-and-paste'' wormholes in disguise; with an unfortunate coordinate choice (pathological at the throat) having the net effect of hiding the thin-shell delta-function layer of curvature (and stress-energy) that is present at the throat. To really make this point clear we have explicitly calculated all the standard curvature tensors (Riemann, Ricci, Einstein, Weyl) for the ``vacuum defect wormhole'', verifying (in this situation) the purely distributional nature of the curvature tensors --- with the curvature tensors being non-zero exactly on the wormhole throat itself.
{For completeness we have explicitly exhibited the coordinate invariant eigenvalues of the mixed Ricci and Riemann tensors.}
For more general ``defect wormholes'' there can additionally be bulk piecewise continuous contributions to the curvature tensors, but within the framework of these ``defect wormholes'' the presence of the distributional contribution on the wormhole throat is unavoidable. 

Perhaps a key observation to make is that one should be very careful with setting up coordinates; in the presence of thin shells an unfortunate choice of coordinates can lead to grossly misleading results.
On a more positive note, once these issues are taken into account the ``defect wormholes'' fall squarely into the mainstream of the extensive body of work on Lorentzian wormholes~\cite{Morris:1988a,Morris:1988b, examples, surgery, book, Frolov:1990, Hayward:1998, Ford:1995, Poisson:1995, Musgrave:1995, Lemos:2003, Lobo:2003, Lobo:2004, Lobo:2004b, Lobo:2005, Sushkov:2005, Lobo:2005b, toolkit, small, Kar:2004,  Sharif:2016, 
Friedman:1993, Hochberg:1991, Roman:1992, Kar:1995, Hochberg:1997, Teo:1998, Hochberg:1998a, Hochberg:1998b, Dadhich:2001, Bronnikov:2002, Bouhmadi-Lopez:2014, Boonserm:2018, Visser:1990,Lobo:2012, Lobo:2015, Curiel:2014, Thorne:2015}.

\bigskip
\hrule\hrule\hrule


\section*{Acknowledgements}

Both JB and RG were supported by  Victoria University of Wellington PhD Doctoral Scholarships.
\\
MV was directly supported by the Marsden Fund, 
via a grant administered by the Royal Society of New Zealand.

\bigskip
\hrule\hrule\hrule
\addtocontents{toc}{\bigskip\hrule}

\null
\vspace{-50pt}
\setcounter{secnumdepth}{0}
\section[\hspace{14pt}  References]{}

\clearpage


\begin{thebibliography}{99}


\bibitem{Klinkhamer:2022}
F.~R.~Klinkhamer,\\
``Defect Wormhole: A Traversable Wormhole Without Exotic Matter'',\\
Acta Phys. Polon. B \textbf{54} (2023) no.5, 5-A3,
\doi{10.5506/APhysPolB.54.5-A3}\\{}
[\arXiv{2301.00724} [gr-qc]].

\bibitem{Klinkhamer:2023a}
F.~R.~Klinkhamer,
``Vacuum defect wormholes and a mirror world'',\\{}
Acta Phys. Pol. B\textbf{54} (2023) no.7, 7-A3,
\doi{10.5506/APhysPolB.54.7-A3}\\{}
[\arXiv{2305.13278} [gr-qc]].

\bibitem{Klinkhamer:2023b}
F.~R.~Klinkhamer,
``New Type of Traversable Wormhole'',\\{}
[\arXiv{2307.04678} [gr-qc]],
To appear in the Bulgarian Journal of  Physics.

\bibitem{Klinkhamer:2023c}
F.~R.~Klinkhamer,
``Higher-dimensional extension of a vacuum-defect wormhole'',\\{}
[\arXiv{2307.12876} [gr-qc]].

\medskip
\hrule\hrule\hrule
\medskip

\bibitem{Wang:2023}
Z.~L.~Wang,
``On a Schwarzschild-type defect wormhole'',
[\arXiv{2307.01678} [gr-qc]].

\bibitem{Ahmed:2023a}
F.~Ahmed,\\
``A topologically charged four-dimensional wormhole and the energy conditions'',\\{}
[\arXiv{2308.00012} [gr-qc]].

\bibitem{Ahmed:2023b}
F.~Ahmed,
``Topologically Charged Rotating Wormhole'',
[\arXiv{2308.03815} [gr-qc]].

\bibitem{Ahmed:2023c}
F.~Ahmed,
``Three-dimensional wormhole with cosmic string effects on eigenvalue solution of non-relativistic quantum particles'',
Sci. Rep. \textbf{13} (2023) no.1, 12953
\doi{10.1038/s41598-023-40066-z}


\bibitem{Ahmed:2023d}
F.~Ahmed,
``Construction of a new five-dimensional vacuum-defect wormhole'',  \\{}
[\arXiv{2308.11938} [gr-qc]].

\medskip
\hrule\hrule\hrule
\medskip



\bibitem{Morris:1988a}
M.~S.~Morris and K.~S.~Thorne,
``Wormholes in space-time and their use for interstellar travel: A tool for teaching general relativity'',\\
Am. J. Phys. \textbf{56} (1988), 395-412\;
\doi{10.1119/1.15620}

\bibitem{Morris:1988b}
M.~S.~Morris, K.~S.~Thorne and U.~Yurtsever,
``Wormholes, Time Machines, and the Weak Energy Condition'',
Phys. Rev. Lett. \textbf{61} (1988), 1446-1449\\
\doi{10.1103/PhysRevLett.61.1446}


\medskip
\hrule\hrule\hrule
\medskip


\enlargethispage{20pt}


\bibitem{Feng:2023}
Justin Christopher Feng,
``Smooth metrics can hide thin shells'',\\
Classical and Quantum Gravity {\bf 40} (2023) 197002,
\doi{10.1088/1361-6382/acf2de}\\{}
[\arXiv{2308.11885} [gr-qc]].

\medskip
\hrule\hrule\hrule
\medskip

\bibitem{Israel}
W.~Israel,
``Singular hypersurfaces and thin shells in general relativity'',\\
Nuovo Cim. B \textbf{44S10} (1966), 1
[erratum: Nuovo Cim. B \textbf{48} (1967), 463]
\doi{10.1007/BF02710419}\

\bibitem{Lanczos:1924}
Kornel Lanczos,\\
``Flächenhafte Verteilung der Materie in der Einsteinschen Gravitationstheorie'',
Annalen der Physik {\bf 379 \# 14} (1924) 518--540 \doi{10.1002/andp.19243791403}

\bibitem{Lanczos:1922}
Kornel  Lanczos,\\
``Untersuchung \"uber fl\"achenhafte Verteilung der Materie in der Einsteinschen Gravitationstheorie'',
1922, unpublished.\\
(Reportedly rejected by Annalen der Physik as ``too mathematically oriented''.)

\enlargethispage{20pt}
\bibitem{Sen}
Nikhil Ranjan Sen,\\
``\"Uber die Grenzbedingungen des Schwerefeldes an Unstetigkeitsfl\"achen'',\\
Annalen der Physik {\bf 378 \# 5-6} (1924) 365--396. 
\doi{10.1002/andp.19243780505}

\medskip
\hrule\hrule\hrule
\medskip

\bibitem{examples}
M.~Visser,
``Traversable wormholes: Some simple examples'',\\
Phys. Rev. D \textbf{39} (1989), 3182-3184
\doi{10.1103/PhysRevD.39.3182}\\{}
[\arXiv{0809.0907} [gr-qc]].

\bibitem{surgery}
M.~Visser,\\
``Traversable wormholes from surgically modified Schwarzschild space-times'',\\
Nucl. Phys. B \textbf{328} (1989), 203-212
\doi{10.1016/0550-3213(89)90100-4}\\{}
[\arXiv{0809.0927} [gr-qc]].

\bibitem{book}
M.~Visser,
``Lorentzian wormholes: From Einstein to Hawking'',\\
AIP press (now Springer), New York, 1995.


\bibitem{Poisson:1995}
E.~Poisson and M.~Visser,
``Thin shell wormholes: Linearization stability'',\\
Phys. Rev. D \textbf{52} (1995), 7318-7321
\doi{10.1103/PhysRevD.52.7318}\\{}
[\arXiv{gr-qc/9506083} [gr-qc]].



\medskip
\hrule\hrule\hrule
\medskip

{
\bibitem{Musgrave:1995}
P.~Musgrave and K.~Lake,\\
``Junctions and thin shells in general relativity using computer algebra. 1: The Darmois--Israel formalism'',\\
Class. Quant. Grav. \textbf{13} (1996), 1885-1900
\doi{10.1088/0264-9381/13/7/018}
[\arXiv{gr-qc/9510052} [gr-qc]].

\bibitem{Eiroa:2003}
E.~F.~Eiroa and G.~E.~Romero,
``Linearized stability of charged thin shell wormholes'',
Gen. Rel. Grav. \textbf{36} (2004), 651-659\\
\doi{10.1023/B:GERG.0000016916.79221.24}
[\arXiv{gr-qc/0303093} [gr-qc]].

\bibitem{Lobo:2003}
F.~S.~N.~Lobo and P.~Crawford,
``Linearized stability analysis of thin shell wormholes with a cosmological constant'',\\
Class. Quant. Grav. \textbf{21} (2004), 391-404
\doi{10.1088/0264-9381/21/2/004}
[\arXiv{gr-qc/0311002} [gr-qc]].

\bibitem{Lobo:2004}
F.~S.~N.~Lobo,
``Energy conditions, traversable wormholes and dust shells'',\\
Gen. Rel. Grav. \textbf{37} (2005), 2023-2038
\doi{10.1007/s10714-005-0177-x}
[\arXiv{gr-qc/0410087} [gr-qc]].

\bibitem{Lobo:2004b}
F.~S.~N.~Lobo,
``Surface stresses on a thin shell surrounding a traversable wormhole'',\\
Class. Quant. Grav. \textbf{21} (2004), 4811-4832
\doi{10.1088/0264-9381/21/21/005}
[\arXiv{gr-qc/0409018} [gr-qc]].


\bibitem{Lobo:2005b}
F.~S.~N.~Lobo and P.~Crawford,
``Stability analysis of dynamic thin shells'',\\
Class. Quant. Grav. \textbf{22} (2005), 4869-4886
\doi{10.1088/0264-9381/22/22/012}
[\arXiv{gr-qc/0507063} [gr-qc]].

}

\medskip
\hrule\hrule\hrule
\medskip


\bibitem{toolkit}
E.~Poisson,
``A Relativist's Toolkit: The Mathematics of Black-Hole Mechanics,''
(Cambridge University Press, 2009, UK) 
\doi{10.1017/CBO9780511606601}


\bibitem{small}
M.~Visser, S.~Kar and N.~Dadhich,\\
``Traversable wormholes with arbitrarily small energy condition violations'',\\
Phys. Rev. Lett. \textbf{90} (2003), 201102
\doi{10.1103/PhysRevLett.90.201102}\\{}
[\arXiv{gr-qc/0301003} [gr-qc]].

\bibitem{Kar:2004}
S.~Kar, N.~Dadhich and M.~Visser,\\
``Quantifying energy condition violations in traversable wormholes'',\\
Pramana \textbf{63} (2004), 859-864
\doi{10.1007/BF02705207}
[\arXiv{gr-qc/0405103} [gr-qc]].

\bibitem{Sharif:2016}
M.~Sharif and F.~Javed,
``On the stability of Bardeen thin-shell wormholes'',\\
Gen. Rel. Grav. \textbf{48} (2016) no.12, 158
\doi{10.1007/s10714-016-2154-y}

\medskip
\hrule\hrule\hrule
\medskip



\bibitem{Raychaudhuri}
A. K. Raychaudhuri,  ``Relativistic cosmology I.'',\\
Phys. Rev. {\bf 98 \# 4} (1955) 1123--1126. \doi{10.1103/PhysRev.98.1123}

\bibitem{Dadhich:2005}
N.~Dadhich,
``Derivation of the Raychaudhuri equation'',\\{}
[\arXiv{gr-qc/0511123} [gr-qc]].

\bibitem{Kar:2006}
S.~Kar and S.~SenGupta,
``The Raychaudhuri equations: A Brief review'',\\
Pramana \textbf{69} (2007), 49
\doi{10.1007/s12043-007-0110-9}
[\arXiv{gr-qc/0611123} [gr-qc]].


\bibitem{Abreu:2010}
G.~Abreu and M.~Visser,
``Some generalizations of the Raychaudhuri equation'',
Phys. Rev. D \textbf{83} (2011), 104016
\doi{10.1103/PhysRevD.83.104016}\\{}
[\arXiv{1012.4806} [gr-qc]].

{
\bibitem{Segre}
Corrado Segre,
``Sulla teoria e sulla classificazione delle omografie in uno spazio lineare ad un numero qualunque di dimensioni'',
Mem. R. Acc. Naz. Lincei, Vol. 19 (1883-84), p. 127–148.
\url{http://www.bdim.eu/item?id=GM_Segre_CW_3_304}

\bibitem{Plebanski}
Jerzy Plebanski, ``The Algebraic structure of the tensor of matter'',
Acta Physica Polonica. 26 (1964) 963.

\bibitem{Santos:2005}
J.~Santos and J.~S.~Alcaniz,
``Energy conditions and Segre classification of phantom fields'',
Phys. Lett. B \textbf{619} (2005), 11-16
\doi{10.1016/j.physletb.2005.05.059}
[\arXiv{astro-ph/0502031} [astro-ph]].

\bibitem{Petrov}
A.~Z.~Petrov, ``Klassifikacya prostranstv opredelyayushchikh polya tyagoteniya",\\
Uch. Zapiski Kazan. Gos. Univ. {\bf 114 \#8} (1954) 55--69. \\
English translation:  A.~Z.~Petrov,  ``Classification of spaces defined by gravitational fields",\\
General Relativity and Gravitation. {\bf 32 \#8}  (2000) 1665--1685.
\doi{10.1023/A:1001910908054}. 

\bibitem{Stephani:2003}
H.~Stephani, D.~Kramer, M.~A.~H.~MacCallum, C.~Hoenselaers and E.~Herlt,\\
``Exact solutions of Einstein's field equations'',
Cambridge Univ. Press, 2003,
ISBN 978-0-521-46702-5, 978-0-511-05917-9
\doi{10.1017/CBO9780511535185}
}


\medskip
\hrule\hrule\hrule
\medskip
\bibitem{Friedman:1993}
J.~L.~Friedman, K.~Schleich and D.~M.~Witt,
``Topological censorship'',\\
Phys. Rev. Lett. \textbf{71} (1993), 1486-1489
[erratum: Phys. Rev. Lett. \textbf{75} (1995), 1872]
\doi{10.1103/PhysRevLett.71.1486}
[\arXiv{gr-qc/9305017} [gr-qc]].

\bibitem{Hochberg:1991}
D.~Hochberg and T.~W.~Kephart,
``Lorentzian wormholes from the gravitationally squeezed vacuum'',
Phys. Lett. B \textbf{268} (1991), 377-383
\doi{10.1016/0370-2693(91)91593-K}

\bibitem{Roman:1992}
T.~A.~Roman,
``Inflating Lorentzian wormholes'',
Phys. Rev. D \textbf{47} (1993), 1370-1379
\doi{10.1103/PhysRevD.47.1370}
[\arXiv{gr-qc/9211012} [gr-qc]].

\bibitem{Kar:1995}
S.~Kar and D.~Sahdev,\\
``Evolving Lorentzian wormholes'',
Phys. Rev. D \textbf{53} (1996), 722-730
\doi{10.1103/PhysRevD.53.722}
[\arXiv{gr-qc/9506094} [gr-qc]].



\bibitem{Hochberg:1997}
D.~Hochberg and M.~Visser,
``Geometric structure of the generic static traversable wormhole throat'',
Phys. Rev. D \textbf{56} (1997), 4745-4755
\doi{10.1103/PhysRevD.56.4745}
[\arXiv{gr-qc/9704082 }[gr-qc]].

\bibitem{Teo:1998}
E.~Teo,
``Rotating traversable wormholes'',
Phys. Rev. D \textbf{58} (1998), 024014
\doi{10.1103/PhysRevD.58.024014}\\{}
[\arXiv{gr-qc/9803098} [gr-qc]].

\bibitem{Hochberg:1998a}
D.~Hochberg and M.~Visser,
``The Null energy condition in dynamic wormholes'',
Phys. Rev. Lett. \textbf{81} (1998), 746-749
\doi{10.1103/PhysRevLett.81.746}
[\arXiv{gr-qc/9802048} [gr-qc]].

{
\bibitem{Ford:1995}
L.~H.~Ford and T.~A.~Roman,
``Quantum field theory constrains traversable wormhole geometries'',\\
Phys. Rev. D \textbf{53} (1996), 5496-5507
\doi{10.1103/PhysRevD.53.5496}
[\arXiv{gr-qc/9510071} [gr-qc]].
}

\bibitem{Hochberg:1998b}
D.~Hochberg and M.~Visser,\\
``Dynamic wormholes, anti-trapped surfaces, and energy conditions'',\\
Phys. Rev. D \textbf{58} (1998), 044021
\doi{10.1103/PhysRevD.58.044021}
[\arXiv{gr-qc/9802046} [gr-qc]].

\bibitem{Dadhich:2001}
N.~Dadhich, S.~Kar, S.~Mukherji and M.~Visser,\\
``$R = 0$ space-times and selfdual Lorentzian wormholes'',\\
Phys. Rev. D \textbf{65} (2002), 064004
\doi{10.1103/PhysRevD.65.064004}
[\arXiv{gr-qc/0109069} [gr-qc]].

{
\bibitem{Bronnikov:2002}
K.~A.~Bronnikov and S.~W.~Kim,
``Possible wormholes in a brane world'',
Phys. Rev. D \textbf{67} (2003), 064027\\
\doi{10.1103/PhysRevD.67.064027}
[\arXiv{gr-qc/0212112} [gr-qc]].
}

\bibitem{Bouhmadi-Lopez:2014}
M.~Bouhmadi-L\'opez, F.~S.~N.~Lobo and P.~Mart\'\i{}n-Moruno,\\
``Wormholes minimally violating the null energy condition'',
JCAP \textbf{11} (2014), 007
\doi{10.1088/1475-7516/2014/11/007}
[\arXiv{1407.7758} [gr-qc]].

\clearpage
\bibitem{Boonserm:2018}
P.~Boonserm, T.~Ngampitipan, A.~Simpson and M.~Visser,\\
``Exponential metric represents a traversable wormhole'',\\
Phys. Rev. D \textbf{98} (2018) no.8, 084048
\doi{10.1103/PhysRevD.98.084048}\\{}
[\arXiv{1805.03781} [gr-qc]].

\bibitem{Visser:1990}
M.~Visser,
``Wheeler wormholes and topology change'',\\
Mod. Phys. Lett. A \textbf{6} (1991), 2663-2668
\doi{10.1142/S0217732391003109}

\bibitem{Lobo:2012}
F.~S.~N.~Lobo, P.~Martin-Moruno, N.~Montelongo-Garcia and M.~Visser,\\
``Linearised stability analysis of generic thin shells'',
\doi{10.1142/9789814623995\_0321}
[\arXiv{1211.0605} [gr-qc]].

\bibitem{Lobo:2015}
F.~S.~N.~Lobo, M.~Bouhmadi-L\'opez, P.~Mart\'\i{}n-Moruno, N.~Montelongo-Garc\'\i{}a and M.~Visser,
``A novel approach to thin-shell wormholes and applications'',
\doi{10.1142/9789813226609\_0154}
[\arXiv{1512.08474} [gr-qc]].


\bibitem{Frolov:1990}
V.~P.~Frolov and I.~D.~Novikov,
``Physical Effects in Wormholes and Time Machine'',
Phys. Rev. D \textbf{42} (1990), 1057-1065\\{}
\doi{10.1103/PhysRevD.42.1057}

\bibitem{Hayward:1998}
S.~A.~Hayward,
``Dynamic wormholes'',
Int. J. Mod. Phys. D \textbf{8} (1999), 373-382
\doi{10.1142/S0218271899000286}\\{}
[\arXiv{gr-qc/9805019} [gr-qc]].


\bibitem{Lemos:2003}
J.~P.~S.~Lemos, F.~S.~N.~Lobo and S.~Quinet de Oliveira,
``Morris-Thorne wormholes with a cosmological constant'',\\
Phys. Rev. D \textbf{68} (2003), 064004
\doi{10.1103/PhysRevD.68.064004}
[\arXiv{gr-qc/0302049} [gr-qc]].


\bibitem{Lobo:2005}
F.~S.~N.~Lobo,
``Phantom energy traversable wormholes'',
Phys. Rev. D \textbf{71} (2005), 084011
\doi{10.1103/PhysRevD.71.084011}
[\arXiv{gr-qc/0502099 [gr-qc}]].

\bibitem{Sushkov:2005}
S.~V.~Sushkov,
``Wormholes supported by a phantom energy'',
Phys. Rev. D \textbf{71} (2005), 043520
\do{10.1103/PhysRevD.71.043520}
[\arXiv{gr-qc/0502084} [gr-qc]].

\bibitem{Curiel:2014}
E.~Curiel,
``A Primer on Energy Conditions'',
Einstein Stud. \textbf{13} (2017), 43-104
\doi{10.1007/978-1-4939-3210-8\_3}\\{}
[\arXiv{1405.0403} [physics.hist-ph]].




\bibitem{Thorne:2015}
O.~James, E.~von Tunzelmann, P.~Franklin and K.~S.~Thorne,\\
``Visualizing Interstellar's Wormhole'',
Am. J. Phys. \textbf{83} (2015), 486
\doi{10.1119/1.4916949}
[\arXiv{1502.03809} [gr-qc]].

\medskip
\hrule\hrule\hrule
\medskip


\end{thebibliography}
\end{document}